# OLS4: A new Ontology Lookup Service for a growing interdisciplinary knowledge ecosystem

James McLaughlin[*]; Josh Lagrimas[*]; Haider Iqbal[*]; Helen Parkinson[*]; Henriette Harmse[*,**]
[*]EMBL-EBI, Hinxton, UK
[**]Corresponding author

## 1. Abstract

**Summary:** The Ontology Lookup Service (OLS) is an open source search engine for ontologies which is used extensively in the bioinformatics and chemistry communities to annotate biological and biomedical data with ontology terms. Recently there has been a significant increase in the size and complexity of ontologies due to new scales of biological knowledge, such as spatial transcriptomics, new ontology development methodologies, and curation on an increased scale. Existing Web-based tools for ontology browsing such as BioPortal and OntoBee do not support the full range of definitions used by today's ontologies. In order to support the community going forward, we have developed OLS4, implementing the complete OWL2 specification, internationalization support for multiple languages, and a new user interface with UX enhancements such as links out to external databases. OLS4 has replaced OLS3 in production at EMBL-EBI and has a backwards compatible API supporting users of OLS3 to transition.

**Availability and Implementation:** Freely available on the web at https://www.ebi.ac.uk/ols4. Source code available at https://github.com/EBISPOT/ols4 with Apache 2.0 License
**Contact:** henriette007@ebi.ac.uk

## 2. Introduction

The Ontology Lookup Service (OLS) is a search engine for ontologies, first released in 2006 [1]. It supports users to search for ontology terms during knowledge curation required by the Findable, Accessible, Interoperable, and Reusable (FAIR) principles. Users of OLS include high-throughput phenotyping centers producing and exporting their data, such as members of the International Mouse Phenotyping Consortium (IMPC) [2]; data integration initiatives such as the OpenTargets Platform for drug target identification and prioritization [3]; and curators of databases including the Genome-Wide Association Study (GWAS) Catalog [4], Expression Atlas [5], European Genome-Phenome Archive [6], Polygenic Score Catalog [7], ChEMBL [8], WormBase [9], EuropePMC [10], PRIDE [11], Ensembl [11], IntAct [12], CancerModels.org [13], BioStudies [14], and the BioImage Archive [15]. Recent applications of OLS include harmonization and standardization of data, for example protocols across phenotyping centers; quality control (QC) checks on raw data, e.g. correction of data submission errors and detection of baseline drift due to instrumentation; and tracking the progress of phenotyping efforts by funding bodies.

Successive iterations of OLS have evolved in response to user needs, for example as standards for APIs changed from SOAP to REST [16] and when new OBO and Web Ontology Language (OWL) version 2 (OWL2) [17] standards for ontologies were introduced [18]. Recently the scale and complexity of biological and chemical knowledge has increased dramatically. New methodologies such as spatial transcriptomics have changed the resolution of data to single cell; for example, OLS is used in the Human BioMolecular Atlas Program (HuBMAP) [19]; and high performance computing has become more abundant. In turn, ontologies have grown significantly in scale: in December 2016, OLS indexed 158 ontologies with 4,862,923 classes. In December 2024, OLS indexed 266 ontologies with 8,682,322 classes. New authoring tools such as ROBOT templates [20] and Dead Simple OWL Design Patterns (DOSDP) [21] have expedited this process by making the development of ontologies more automated enabling new terms to be added in large quantities. The complexity of ontologies has also increased; for example, internationalization to support the translation of ontologies into different languages to support a diverse and international user base [22], and features from the OWL2 specification such as disjointness statements and property chains. So far none of the existing open-source solutions (OLS3, BioPortal [23], OntoBee [24], and AgroPortal [25]) are able to comprehensively support these use cases.

## 3. Results

We have developed a new version of the Ontology Lookup Service, OLS4, with many new features including full implementation of the OWL2 specification; annotations on annotations; internationalization support; cross-references between ontology terms; and BioRegistry [26] integration. The OWL2 specification has been implemented comprehensively and tested using a suite of test cases based on both the OWL2 Primer and example test-cases extracted from biological and biomedical ontologies, including the Experimental Factor Ontology (EFO) [27] and the MONDO Disease Ontology [28]. OLS4 is therefore able to support ontologies using OWL2 features; for example, in the MONDO disease ontology where OWL2 disjointness is asserted between extrapulmonary tuberculosis (`MONDO:0000368`) versus pulmonary tuberculosis (`MONDO:0006052`); and in the Relation Ontology (RO) where the OWL2 property chain `regulates = directly regulates -> directly regulates` is defined to describe transitive regulation relations. These OWL2 definitions are now visible in the OLS browser and API. In addition to OWL2 ontologies, OLS4 loads schemas defined using `rdfs:Class` hierarchies, providing users with a standard API to access both OWL ontologies and commonly used schemas such as Dublin Core [29] and Schema.org [30], which are in turn used by OWL2 ontologies.

OLS4 improves annotation support by implementing annotations on annotations (sometimes termed reification), making references and provenance associated with ontology axioms visible in the web interface. For example, UBERON [31] is a widely used multi species anatomy ontology. The UBERON term for "lung" contains homology notes derived from *The evolution of organ systems [32]*, a link to which is now visible in the corresponding OLS page for attribution and cross referencing. Full internationalization of annotations has also been added in OLS4; ontologies are browsable in multiple languages and OLS4 displays a language picker listing all languages present in the ontology. When a language is selected, annotations are displayed in the language selected by the user where possible. This functionality has been demonstrated in the Human Phenotype Ontology (HPO) [22] which is now accessible in multiple languages in the main OLS instance, enabling curators to map to consistent phenotype terms across language barriers.

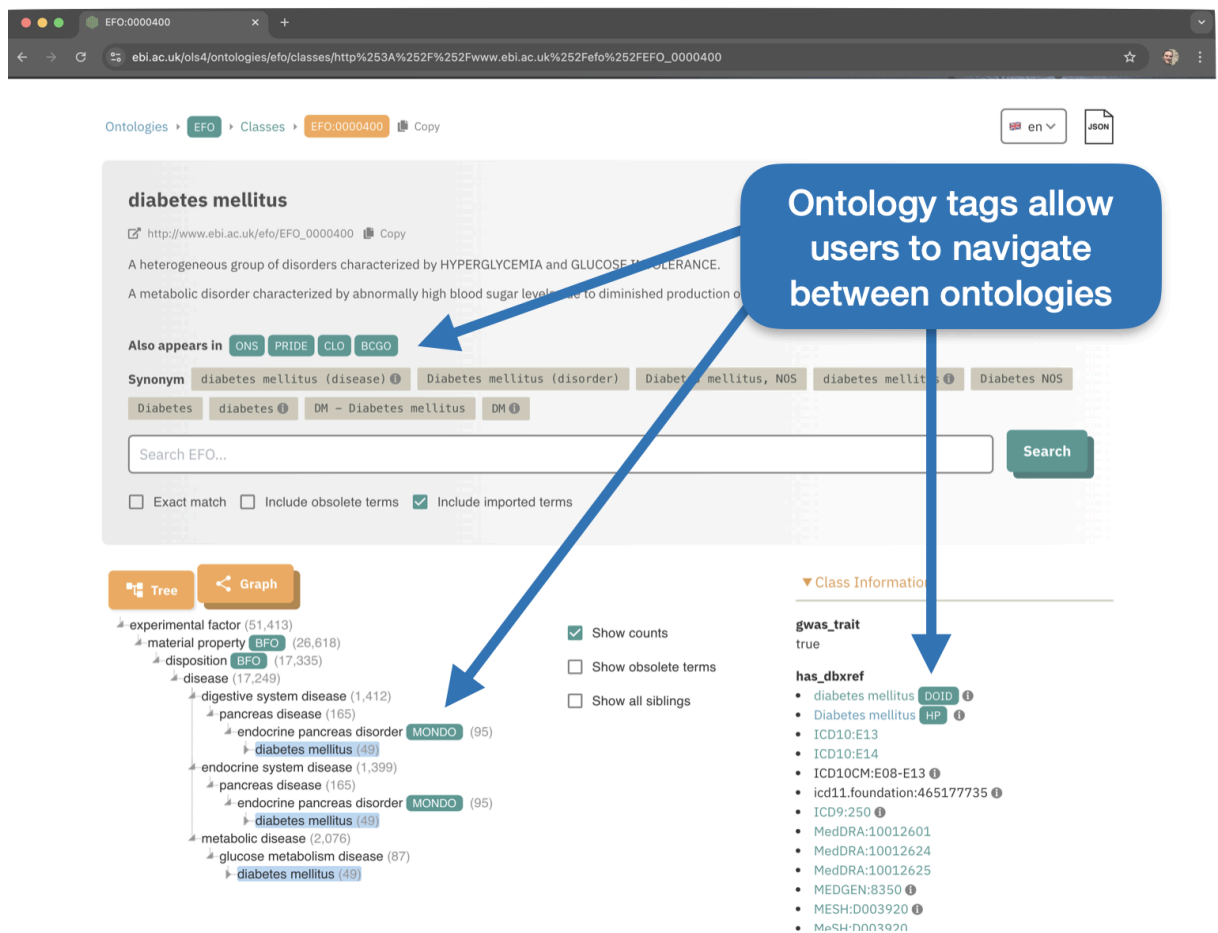

**Fig. 1: An ontology term from the Experimental Factor Ontology (EFO) viewed in OLS4. References to terms in different ontologies are tagged with links to the corresponding defining ontologies for easy navigation.**

Another significant development in OLS4 is the handling of cross-references between ontology terms. The majority of the ontologies indexed by OLS do not exist in isolation, but reuse terms from other ontologies. For example, the Experimental Factor Ontology (EFO) imports chemical terms from the Chemical Entities of Biological Interest (ChEBI) ontology. In OLS4, such imported terms are labelled in the tree with a tag linking to the defining ontology (Fig. 1). This functionality is critical to support the Unified Phenotype Ontology (uPheno) [33], which aggregates terms from multiple phenotype ontologies often with the same name; without the defining ontology tags it would be unclear the difference between e.g. HPO and MP [34] terms would be unclear in the tree view. In addition to cross-references between ontology terms, OLS4 also automatically creates external links using the Bioregistry, enabling users to easily navigate from ontologies to external databases, for example from a gene to a genome sequence in GenBank.

Altogether these features allow OLS4 to deliver a range of new use cases for the ontology community, and to make ontologies more interoperable with other biological and chemical resources. OLS4 is now in production at EMBL-EBI and served approximately 50 million requests from approximately 200,000 unique hosts between Nov 2023 and Feb 2024.

## 4. Methods

The OLS4 data-load and backend are implemented in Java 11 and Spring Boot. OLS4 uses Neo4j as a graph database and Solr for full text search. The architecture has been simplified from OLS3; Neo4J is now used as a standalone server rather than an embedded database, and ontology metadata is also stored in Solr removing the need for a MongoDB instance. The motivation for this simplification was to reduce the complexity of deploying OLS outside of its primary instance at EMBL-EBI, for use cases such as the MONARCH Initiative OLS and the NFDI4Chem Terminology Service [35]. The ETL pipeline has also been simplified by eliminating redundant processing to improve scaling of the dataload. OLS4 data loads for all ontologies are at least 15 times faster, with testing at EBI showing the lower bound OLS3 time as 1 month and the upper bound OLS4 time as 2 days. These faster data loads allow users to see updates to ontologies more quickly and reflect changes in knowledge, particularly important when biological knowledge develops rapidly which is an important issue for pandemic preparedness; OLS is currently being used as part of the European Viral Outbreak Response Alliance (EVORA) project.

In OLS4 the Neo4j and Solr schemas are dynamic and depend on the annotation properties used in the OWL entities in the source ontologies. OWL entities are translated from RDF to a lossless JSON representation, which is then stored complete and unmodified in both Neo4j and Solr alongside the extracted queryable properties. Queries to Neo4j and Solr include this JSON representation which is used to generate API responses and the frontend pages. In order to maintain backwards compatibility, the API is implemented using view classes which match the previous OLS3 data model, but abstract from the underlying OLS4 data model. The use of abstract views enables multiple API versions to be built over the same underlying data model, and changes to the API without reloading data each time to deliver updates delivering new API use cases more quickly to users.

The OLS4 frontend has been rewritten to communicate with the backend exclusively using HTTP APIs rather than by directly accessing the internal data model, offering an option to build new front ends for an OLS4 backend instance, such as third party interfaces tailed to specific ontologies/use cases. Custom instances of OLS3 such as the NFDI4Chem terminology service, which previously had to run divergent OLS instances with locally modified backend code, will in future be able to use the latest OLS backend code coupled to a customized frontend reducing the overhead of keeping the backend code synchronized.

## 5. Discussion

OLS serves curators, annotators, data resource producers, and ontology developers. It has been designed to meet the needs of these user groups as well as to scale for larger and more complex ontologies. For curators and annotators, OLS4 provides a richer view of terms including complete OWL2 axiomatization to help users select appropriate terms and navigate between terms, a unique feature of OLS4 among open source ontology browsers. OLS4 also adds multiple language support which enables bio-curators to search for terms using labels in their native languages, a feature so far only supported by AgroPortal but useful for human disease communities who also need common names. For data resource producers, the faster data load will enable resources to load and link to the latest versions of ontology terms in days without waiting several weeks for OLS to update. For ontology developers, the new display features of OLS will enable new use cases to be delivered in ontologies with visibility to users, as has been demonstrated for HPO and MP internationalized editions. OLS4 will also help to prevent the proliferation of terms across multiple ontologies re-defining the same concepts in different contexts, by adding ontology tags to make the presentation of the relationship between ontologies more prominent and easier to navigate.

## 5.1. Future work

Future work will include support for the Simple Standard for Sharing Ontological Mappings (SSSOM) [36]. Mappings between ontology terms are used e.g. to map phenotype terms between human phenotypes and model organisms such as mouse and zebrafish. While some mappings are present in ontologies, often represented as `hasDbXref` properties, SSSOM allows multiple different mapping sets to be defined with associated mapping metadata, which is important as mappings are often subjective and project-dependent. We plan to add support to load and display alternative sets of mappings depending on user preference, for example to allow users to choose between different HPO to MP mappings provided by MGI, IMPC, and Pistoia Alliance.

In future we also plan to implement more sophisticated search capabilities, such as searching for a specific annotation with a specific value and searching in a specific branch of an ontology. For example, EFO terms used to annotate studies in the GWAS Catalog are annotated with a property `gwas trait = true`. Searching for terms with this annotation would allow the impact of deprecating or moving a term on annotated datasets to be assessed. Limiting a search to a specific branch of an ontology would allow users interested in e.g. cardiology to limit their searches to terms underneath "heart disease" in MONDO or "heart" in UBERON.

## 6. Acknowledgements

The authors would like to thank Nicolas Matentzoglu (Semanticly Ltd), David Osumi-Sutherland (Wellcome Trust Sanger Institute), and the OBO Foundry community for testing and feedback. J.A.M., J.L., H.I., H.P., and H.H. are supported in part by EMBL-EBI Core Funds. J.A.M. was supported by the Chan–Zuckerberg Initiative award for the Human Cell Atlas Data Coordination Platform from 2020 to 2023; Office of the Director, National Institutes of Health (R24-OD011883, OT2OD033756); and NIH National Human Genome Research Institute Phenomics First Resource, NIH-NHGRI # 5RM1 HG010860, a Center of Excellence in Genomic Science. H.H. was supported by the European Union's Horizon 2020 research, and innovation programme grant numbers 824087 (EOSC-Life from Jun 2020 to Aug 2023) and 825575 (EJP-RD from Jun 2020 to Dec 2023).

## 7. References

1. Côté RG, Jones P, Apweiler R, Hermjakob H. The Ontology Lookup Service, a lightweight cross-platform tool for controlled vocabulary queries. BMC Bioinformatics. 2006;7: 97. doi:10.1186/1471-2105-7-97

2. Groza T, Gomez FL, Mashhadi HH, Muñoz-Fuentes V, Gunes O, Wilson R, et al. The International Mouse Phenotyping Consortium: comprehensive knockout phenotyping underpinning the study of human disease. Nucleic Acids Res. 2023;51: D1038–D1045. doi:10.1093/nar/gkac972

3. Ochoa D, Hercules A, Carmona M, Suveges D, Baker J, Malangone C, et al. The next-generation Open Targets Platform: reimagined, redesigned, rebuilt. Nucleic Acids Res. 2023;51: D1353–D1359. doi:10.1093/nar/gkac1046

4. Cerezo M, Sollis E, Ji Y, Lewis E, Abid A, Bircan KO, et al. The NHGRI-EBI GWAS Catalog: standards for reusability, sustainability and diversity. Nucleic Acids Res. 2025;53: D998–D1005. doi:10.1093/nar/gkae1070

5. Moreno P, Fexova S, George N, Manning JR, Miao Z, Mohammed S, et al. Expression Atlas update: gene and protein expression in multiple species. Nucleic Acids Res. 2022;50: D129–D140. doi:10.1093/nar/gkab1030

6. Freeberg MA, Fromont LA, D'Altri T, Romero AF, Ciges JI, Jene A, et al. The European Genome-phenome archive in 2021. Nucleic Acids Res. 2022;50: D980–D987. doi:10.1093/nar/gkab1059

7. Lambert SA, Gil L, Jupp S, Ritchie S, Xu Y, Buniello A, et al. The Polygenic Score Catalog as an open database for reproducibility and systematic evaluation. Nat Genet. 2021;53: 420–425. doi:10.1038/s41588-021-00783-5

8. Gaulton A, Bellis LJ, Bento AP, Chambers J, Davies M, Hersey A, et al. ChEMBL: a large-scale bioactivity database for drug discovery. Nucleic Acids Res. 2012;40: D1100–7. doi:10.1093/nar/gkr777

9. Harris TW, Antoshechkin I, Bieri T, Blasiar D, Chan J, Chen WJ, et al. WormBase: a comprehensive resource for nematode research. Nucleic Acids Res. 2010;38: D463–7. doi:10.1093/nar/gkp952

10. Gou Y, Graff F, Rossiter P, Talo' F, Vartak V, Coleman L, et al. Europe PMC: a full-text literature database for the life sciences and platform for innovation. Nucleic Acids Research. 2014;43: D1042–D1048. doi:10.1093/nar/gku1061

11. Perez-Riverol Y, Bandla C, Kundu DJ, Kamatchinathan S, Bai J, Hewapathirana S, et al. The PRIDE database at 20 years: 2025 update. Nucleic Acids Res. 2025;53: D543–D553. doi:10.1093/nar/gkae1011

12. Hermjakob H, Montecchi-Palazzi L, Lewington C, Mudali S, Kerrien S, Orchard S, et al. IntAct: an open source molecular interaction database. Nucleic Acids Res. 2004;32: D452–5. doi:10.1093/nar/gkh052

13. Perova Z, Martínez M, Mandloi T, Almanza MR, Neuhauser SB, Degley D, et al. Abstract 6910: CancerModels.org: An open global cancer research platform for patient-derived cancer models. Cancer Res. 2024. doi:10.1158/1538-7445.am2024-6910

14. Sarkans U, Gostev M, Athar A, Behrangi E, Melnichuk O, Ali A, et al. The BioStudies database-one stop shop for all data supporting a life sciences study. Nucleic Acids Res. 2018;46: D1266–D1270. doi:10.1093/nar/gkx965

15. Hartley M, Kleywegt GJ, Patwardhan A, Sarkans U, Swedlow JR, Brazma A. The BioImage Archive - building a home for life-sciences microscopy data. J Mol Biol. 2022;434: 167505. doi:10.1016/j.jmb.2022.167505

16. Côté R, Reisinger F, Martens L, Barsnes H, Vizcaino JA, Hermjakob H. The Ontology Lookup Service: bigger and better. Nucleic Acids Res. 2010;38: W155–60. doi:10.1093/nar/gkq331

17. Golbreich C, Wallace EK, Patel-Schneider P. OWL 2 Web Ontology Language new features and rationale. 2009. Available: https://www.w3.org/2007/OWL/draft/ED-owl2-new-features-20081202/all.pdf

18. Jupp S, Burdett T, Leroy C, Parkinson HE. A new Ontology Lookup Service at EMBL-EBI. SWAT4LS. 2015;2: 118–119.

19. Börner K, Blood PD, Silverstein JC, Ruffalo M, Satija R, Teichmann SA, et al. Human BioMolecular Atlas Program (HuBMAP): 3D Human Reference Atlas construction and usage. bioRxivorg. 2024. doi:10.1101/2024.03.27.587041

20. Jackson RC, Balhoff JP, Douglass E, Harris NL, Mungall CJ, Overton JA. ROBOT: A tool for automating ontology workflows. BMC Bioinformatics. 2019;20: 407. doi:10.1186/s12859-019-3002-3

21. Osumi-Sutherland D, Courtot M, Balhoff JP, Mungall C. Dead simple OWL design patterns. J Biomed Semantics. 2017;8: 18. doi:10.1186/s13326-017-0126-0

22. Gargano MA, Matentzoglu N, Coleman B, Addo-Lartey EB, Anagnostopoulos AV, Anderton J, et al. The Human Phenotype Ontology in 2024: phenotypes around the world. Nucleic Acids Res. 2024;52: D1333–D1346. doi:10.1093/nar/gkad1005

23. Noy NF, Shah NH, Whetzel PL, Dai B, Dorf M, Griffith N, et al. BioPortal: ontologies and integrated data resources at the click of a mouse. Nucleic Acids Res. 2009;37: W170–3. doi:10.1093/nar/gkp440

24. Ong E, Xiang Z, Zhao B, Liu Y, Lin Y, Zheng J, et al. Ontobee: A linked ontology data server to support ontology term dereferencing, linkage, query and integration. Nucleic Acids Res. 2017;45: D347–D352. doi:10.1093/nar/gkw918

25. Jonquet C, Toulet A, Arnaud E, Aubin S, Dzalé Yeumo E, Emonet V, et al. AgroPortal: A vocabulary and ontology repository for agronomy. Comput Electron Agric. 2018;144: 126–143. doi:10.1016/j.compag.2017.10.012

26. Hoyt CT, Balk M, Callahan TJ, Domingo-Fernández D, Haendel MA, Hegde HB, et al. Unifying the identification of biomedical entities with the Bioregistry. Sci Data. 2022;9: 714. doi:10.1038/s41597-022-01807-3

27. Malone J, Adamusiak T, Holloway E, Parkinson H. Developing an application ontology for annotation of experimental variables – Experimental Factor Ontology. Nature Precedings. 2009. doi:10.1038/NPRE.2009.3806.1

28. Vasilevsky NA, Matentzoglu NA, Toro S, Flack JE IV, Hegde H, Unni DR, et al. Mondo: Unifying diseases for the world, by the world. bioRxiv. 2022. doi:10.1101/2022.04.13.22273750

29. Weibel S, Kunze J, Lagoze C, Wolf M. Dublin core metadata for resource discovery. 1998. Available: https://www.rfc-editor.org/rfc/rfc2413

30. Guha RV, Brickley D, Macbeth S. Schema.org: evolution of structured data on the web. Commun ACM. 2016;59: 44–51. doi:10.1145/2844544

31. Mungall CJ, Torniai C, Gkoutos GV, Lewis SE, Haendel MA. Uberon, an integrative multi-species anatomy ontology. Genome Biol. 2012;13: R5. doi:10.1186/gb-2012-13-1-r5

32. Schmidt-Rhaesa A. The Evolution of Organ Systems. London, England: Oxford University Press; 2007. Available: https://books.google.com/books?hl=en&lr=&id=OwpREAAAQBAJ&oi=fnd&pg=PR7&dq=+The+evolution+of+organ+systems&ots=4pw09oWIb6&sig=jc6AM61iBznRVkH4ZfS7uGz8uHw

33. Matentzoglu N, Bello SM, Stefancsik R, Alghamdi SM, Anagnostopoulos AV, Balhoff JP, et al. The Unified Phenotype Ontology (uPheno): A framework for cross-species integrative phenomics. bioRxivorg. 2024. doi:10.1101/2024.09.18.613276

34. Smith CL, Goldsmith C-AW, Eppig JT. The Mammalian Phenotype Ontology as a tool for annotating, analyzing and comparing phenotypic information. Genome Biol. 2005;6: R7. doi:10.1186/gb-2004-6-1-r7

35. Steinbeck C, Koepler O, Herres-Pawlis S, Bach F, Jung N, Razum M, et al. NFDI4Chem—A research data network for international chemistry. Chem Int. 2023;45: 8–13. doi:10.1515/ci-2023-0103

36. Matentzoglu N, Balhoff J, Bello S, Bizon C, Brush MH, Callahan T, et al. A Simple Standard for Sharing Ontological Mappings (SSSOM). Database: The Journal of Biological Databases and Curation. 2021;2022. doi:10.1093/database/baac035